\documentclass[conference]{IEEEtran}
\IEEEoverridecommandlockouts
\usepackage{cite}
\usepackage{amsmath,amssymb,amsfonts}
\usepackage[algo2e]{algorithm2e} 
\usepackage{graphicx}
\usepackage{textcomp}
\usepackage{xcolor}
\usepackage[hyphens]{url} 
\usepackage{graphicx} 
\usepackage{subfigure}
\usepackage{enumerate}
\usepackage{microtype}

\usepackage{xcolor}
\usepackage[normalem]{ulem}

\def\BibTeX{{\rm B\kern-.05em{\sc i\kern-.025em b}\kern-.08em
    T\kern-.1667em\lower.7ex\hbox{E}\kern-.125emX}}

\usepackage{epsfig}
\usepackage{algorithm}

\newcommand{\mysection}[1]{\vspace{-.03in}\section{#1}\vspace{-.00in}}
\newcommand{\mysubsection}[1]{\vspace{-.05in}\subsection{#1}\vspace{-.01in}}

\newcommand{\BULLET}{\vspace{+.05in} \noindent $\bullet$ \hspace{+.00in}}

\newcommand{\remove}[1]{}

\newtheorem{lemma}{Lemma}
\newtheorem{corollary}{Corollary}

\newcommand{\ls}[1]
   {\dimen0=\fontdimen6\the\font
    \lineskip=#1\dimen0
    \advance\lineskip.5\fontdimen5\the\font
    \advance\lineskip-\dimen0
    \lineskiplimit=.9\lineskip
    \baselineskip=\lineskip
    \advance\baselineskip\dimen0
    \normallineskip\lineskip
    \normallineskiplimit\lineskiplimit
    \normalbaselineskip\baselineskip
    \ignorespaces
   }

\begin{document}

\title{Relational Deep Reinforcement Learning for Routing in Wireless Networks}

\author{
\IEEEauthorblockN{Victoria Manfredi}
\textit{Wesleyan University}\\
Middletown, CT, USA \\
vumanfredi@wesleyan.edu
\and
\IEEEauthorblockN{Alicia Wolfe}
\textit{Wesleyan University}\\
Middletown, CT, USA \\
pwolfe@wesleyan.edu
\and
\IEEEauthorblockN{Bing Wang}
\textit{University of Connecticut}\\
Storrs, CT, USA \\
bing@uconn.edu
\and
\IEEEauthorblockN{Xiaolan Zhang}
\textit{Fordham University}\\
Bronx, NY, USA \\
xzhang@fordham.edu
}

\maketitle

\begin{abstract}
While routing in wireless networks has been studied extensively, existing protocols are typically designed for a specific set of network conditions and so cannot accommodate any drastic changes in those conditions. For instance, protocols designed for connected networks cannot be easily applied to disconnected networks. 
In this paper, we develop a distributed routing strategy based on deep reinforcement learning that generalizes to diverse  traffic patterns, congestion levels, network connectivity, and link dynamics. 
We make the following key innovations in our design:
(i) the use of {\em relational features} as inputs to the deep neural network approximating the decision space, which enables our algorithm to generalize to diverse network conditions, 
(ii) the use of {\em packet-centric decisions} to transform the routing problem into an episodic task by viewing {\em packets}, rather than wireless devices, as reinforcement learning agents, which provides a natural way to propagate and model rewards accurately during learning, and
(iii) the use of {\em extended-time actions} to model the time  spent  by  a  packet  waiting  in  a  queue, which reduces  the amount of  training data needed and allows the learning algorithm to converge more quickly.
We evaluate our routing algorithm using a packet-level simulator and show that the policy our algorithm learns during training is able to generalize to larger and more congested networks, different topologies, and diverse link dynamics. 
Our algorithm outperforms shortest path and backpressure routing with respect to packets delivered and delay per packet.
\end{abstract}


\vspace{-0.1in}
\mysection{Introduction}
\label{sec:introduction}

The problem of routing in wireless networks is much more challenging than that in wired networks: the shared nature of the wireless medium  reduces per-device bandwidth, while the variability of wireless signal propagation and device mobility introduce topology uncertainty.
While many routing algorithms have been developed for wireless networks (see \S\ref{sec:related-work}),
they typically assume operation under very specific network conditions. 
For example, routing algorithms developed for ad hoc networks assume an always connected network, with a focus on finding an optimal path (e.g., in terms of latency) from a source to a destination; conversely, routing strategies for delay or disruption tolerant networks assume a mostly disconnected network, and hence the focus is to determine, after encountering another device, whether to forward a packet to that device, so as to 
optimize some performance criteria (e.g., the latency until the packet is  delivered to the destination). 

In this paper, we ask the following question: {\em Can we design a generalizable routing algorithm that seamlessly adapts to different network conditions?} In other words, we seek to design a routing algorithm that works regardless of traffic pattern, congestion level, network connectivity, or link dynamics. 
Such a strategy is desirable since the optimal routing algorithm can be very different depending on the network conditions.
For example, consider a wireless network that is initially connected and then becomes disconnected due to device failures or link dynamics.
In this scenario, a routing algorithm whose goal is to find a connected end-to-end path (e.g., through route discovery requests as in DSR~\cite{Johnson1996:DSR} or AODV~\cite{Perkins1999:AODV}) will initially succeed, but will eventually fail to deliver any packets when the network becomes disconnected.  
Alternatively, consider a wireless network that fluctuates between periods of low and high congestion. In this scenario, the use of 
backpressure routing~\cite{tassiulas1990stability} can achieve
optimal performance during the high congestion periods, but leads to  performance inferior to many other algorithms during the low congestion periods. 

One way to design an adaptive routing algorithm is to identify a set of target network conditions to handle, identify the appropriate routing strategy for each, and then switch strategies as needed. This approach risks instability if network conditions change frequently or algorithms take a long time to converge.
Reinforcement learning (RL) \cite{sutton2018reinforcement} allows for an alternate approach in which an RL agent trained on a set of target network conditions learns to make routing decisions even in an uncertain and time-varying environment. 
Q-routing~\cite{Boyan94:Q-routing} is the first RL-based routing algorithm. Since then, many more  have been proposed, see \cite{Mammeri19:RL} and the references within. 
Recently, advances in deep reinforcement learning (DRL), which uses neural networks to approximate the decision space, have motivated the design of DRL-based routing
algorithms (see \S\ref{sec:related-work}). However, existing RL and DRL-based routing strategies do not easily generalize to other scenarios because they encode assumptions about network topology and possible actions
into the neural network that is used to make routing decisions. 

In this work, we develop a novel DRL-based routing strategy that is able to generalize to different network conditions. We focus on distributed routing, allowing individual agents to make routing decisions: this supports scalability and provides redundancy in the case of network dynamics.
We make the following contributions to DRL-based routing design:

\BULLET {\em Relational features.} To enable our algorithm to scale to larger  networks and generalize to other network conditions, we input relational features to the deep neural network used to approximate the routing decision space. This not only allows data from all DRL agents to be used to train the neural network, but also allows each agent to  independently  use  the  same  neural network   for decision-making during testing.
\vspace{-0.025in}

\BULLET {\em Packet-centric decisions.} We transform the routing problem into an episodic task by viewing {\em packets}, rather than devices, as the DRL agents that must learn a routing policy.
This packet-centric approach provides a more natural way to propagate and model rewards accurately. 

\BULLET {\em Extended-time actions.} We use extended-time actions or options \cite{sutton1999between} to model the time  spent  by  a  packet  waiting  in  a  queue, which reduces  the amount of  training data needed and allows the learning algorithm to converge more quickly.

We evaluate our approach using a packet-level network simulator. Extensive results demonstrate that our approach is both generalizable and scalable, and significantly outperforms  shortest path routing \cite{Bellman58:routing} and backpressure routing \cite{tassiulas1990stability} with respect to  packets  delivered  and  delay.
The rest of this paper is organized as follows. \S\ref{sec:routing} gives background on routing, \S\ref{sec:model} overviews  our DRL algorithm, \S\ref{sec:evaluation} provides simulation results, \S\ref{sec:related-work} describes related work, and \S\ref{sec:conclusions} gives our conclusions.
 
\mysection{Routing in Multi-hop Wireless Networks}
\label{sec:routing}

Consider a wireless network with a set of nodes, i.e., wireless devices, $V$. Let $N=|V|$ denote the number of devices in the network. 
Each device transmits via a wireless channel. Two devices that are within transmission range can communicate with each other. 
Let $E_t$ denote the set of wireless links present at time $t$. Due to  interference and possible link dynamics, $E_t$ can vary over time. We assume that devices are stationary; device mobility leads to other interesting challenges, which we leave to future work. 
All devices are capable of receiving and forwarding packets as well as serving as a source or destination. 
Each device  $v \in V$ has a finite queue which can buffer a maximum of $B$ packets. The routing decision at a device $v$ is to choose the neighbor to which to send a packet from $v$'s queue (typically the packet at the front of the queue). A packet carries a time-to-live (TTL) field, which is decremented by one at each intermediate device 
that forwards the packet. A packet is forwarded until it either reaches its destination, its TTL becomes zero, or it is dropped upon arrival at a device whose queue is full.

Two goals of routing are to (i) maximize throughput, i.e., the packet delivery rate, while (ii) minimizing delay, i.e., the time 
from a packet being generated at the source to  being delivered to its destination.
When a network's traffic load changes, or topology changes in the network itself cause traffic load changes, 
a static routing strategy that ignores congestion can lead to poor performance. For instance, shortest path routing, which selects the path 
between a source and destination solely based on the number of hops, leads to low throughput when the shortest path is congested. 
In contrast, an adaptive routing algorithm typically leads to either high throughput with large delay, or  low throughput with small delay. One example is backpressure routing~\cite{tassiulas1990stability}, which routes packets dynamically based on the amount of congestion in the network. 
As we shall see in \S\ref{sec:evaluation}, when the per device queue size, $B$, is 
large, it leads to 
high throughput at the cost of large delay, since packets flow through the network following the lowest congestion gradient at each device, and may take many hops to reach the  destination. When $B$ is small, it leads to lower delay at the cost of low throughput since many packets are dropped. 
In general, it is difficult to achieve both high throughput and low latency simultaneously under dynamic traffic conditions.

\mysection{Related Work}
\label{sec:related-work}

{\em Routing in wireless networks.} The literature on routing strategies for wireless networks is extensive, ranging from earlier protocols such as DSR~\cite{Johnson1996:DSR} and AODV~\cite{Perkins1999:AODV} for ad-hoc networks, various routing protocols for delay and disruption tolerant networks (DTNs)~\cite{Jain2004:DTN}, to a plethora of strategies for resource constrained wireless networks (such as sensor networks or IoT networks)~\cite{Li2011:sensor}. Most existing routing protocols, however, are designed for a specific wireless network scenario and so do not generalize well to other  scenarios. For instance, routing protocols for ad hoc networks assume a connected network, while  routing protocols for DTNs assume a disconnected network. The DRL-based routing strategy proposed in our work is able to generalize to different wireless scenarios.

{\em RL and DRL-based routing.} The literature on RL-based routing is extensive, starting with~\cite{Boyan94:Q-routing}, with improvements in~\cite{Choi96:PQ-routing,Kumar98:Confidence}
 and schemes for different types of networks~\cite{Al-Rawi15:RL,Mammeri19:RL}. The works closest to ours are those that design DRL-based routing strategies. Broadly, these studies are for two settings: centralized and distributed routing. In centralized routing, one agent makes decisions for all routers using one deep neural network (DNN), often for traffic engineering or software defined networks~\cite{Stampa17:DRL,Valadarsky17:learning,Pham18:DRL,Xu2018:experience,Suarez-Varela19:feature,Sun19:SINET}.  
In distributed routing, each agent makes decisions independently using its own DNN. 
However, existing work on distributed routing using DRL is inconclusive or unscalable or does not handle network dynamics.
For instance~\cite{Mukhutdinov19:Multi-agent} claims that experience from different agents is incompatible and leads to ineffective learning. In~\cite{You19:toward} 
an individual DNN is trained for each device, using device specific information such as the last $k$ actions taken and the next $m$ destinations of packets in queues as input,
which limits the scalability of the approach (the largest network that they test only contains 25 devices and 56 links). 

While we also focus on distributed routing, 
our study is the first that  applies relational DRL to this problem. The use of relational features allows our approach to scale to much larger networks (e.g., 100 devices and $>$100 links) than existing studies,
 by removing the need to use device IDs in training (such as for choosing a packet's next hop).
 As a consequence, we are able to train a DNN using data from all agents, as well as allow each DRL agent to independently make decisions using the same DNN.
Our packet-centric approach also allows us to transform the problem into an episodic task, and our use of extended-time actions allows for faster learning.

{\em GNN-based routing.}
Our use of relational features, in particular how we aggregate neighbor features, is similar in spirit to the aggregation function in a graph neural network (GNN) \cite{google-gnn}.
Several studies leverage the generalization capability of GNNs for routing so that the learned strategies are generalizable to other topology and traffic intensity. The study~\cite{Geyer18:GNN} relies on supervised learning. The studies in~\cite{Rusek19:Unveiling,Almasan20:GNN} combine DRL and GNNs for centralized routing; designing such models for a distributed setting is much more challenging.

\begin{figure*}[t]
   \centerline{
   \subfigure[Packets can have different destinations, $D_1$ and $D_2$ respectively, but still use the same relational features, destination distance and queue length, just with different values.]{\hbox{\includegraphics[width=2.in, trim = 6cm 5cm 2cm 2.5cm, clip]{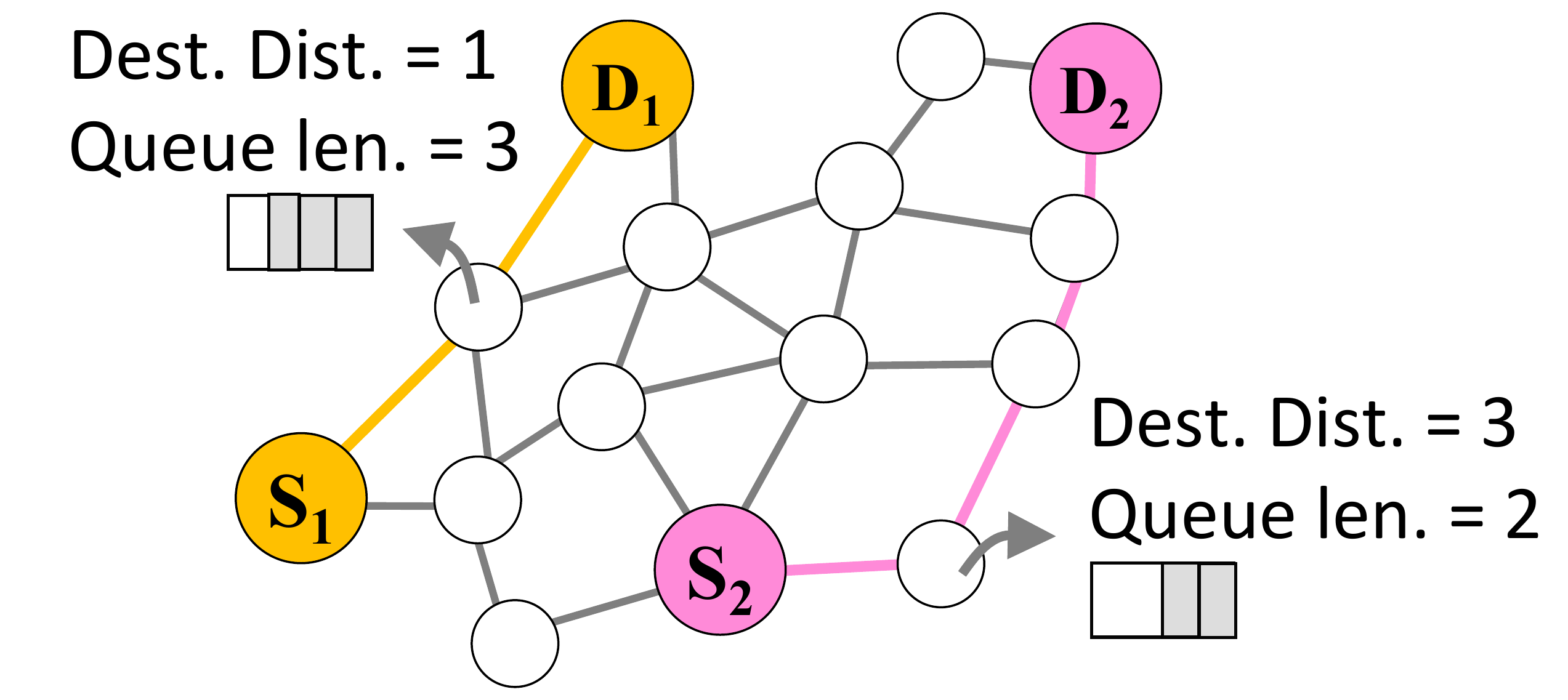}}}
   \hspace{0.1in}
   \subfigure[Using a DNN to make routing decisions. Packet $p$ is considering the action of moving from device $v$ to  $u$ at time $t$, so feeds the associated features into the DNN to obtain the $Q$-value of action $u$.  ]{\hbox{\includegraphics[width=2.5in, trim = 0.32cm 0cm 2.5cm 4.5cm, clip]{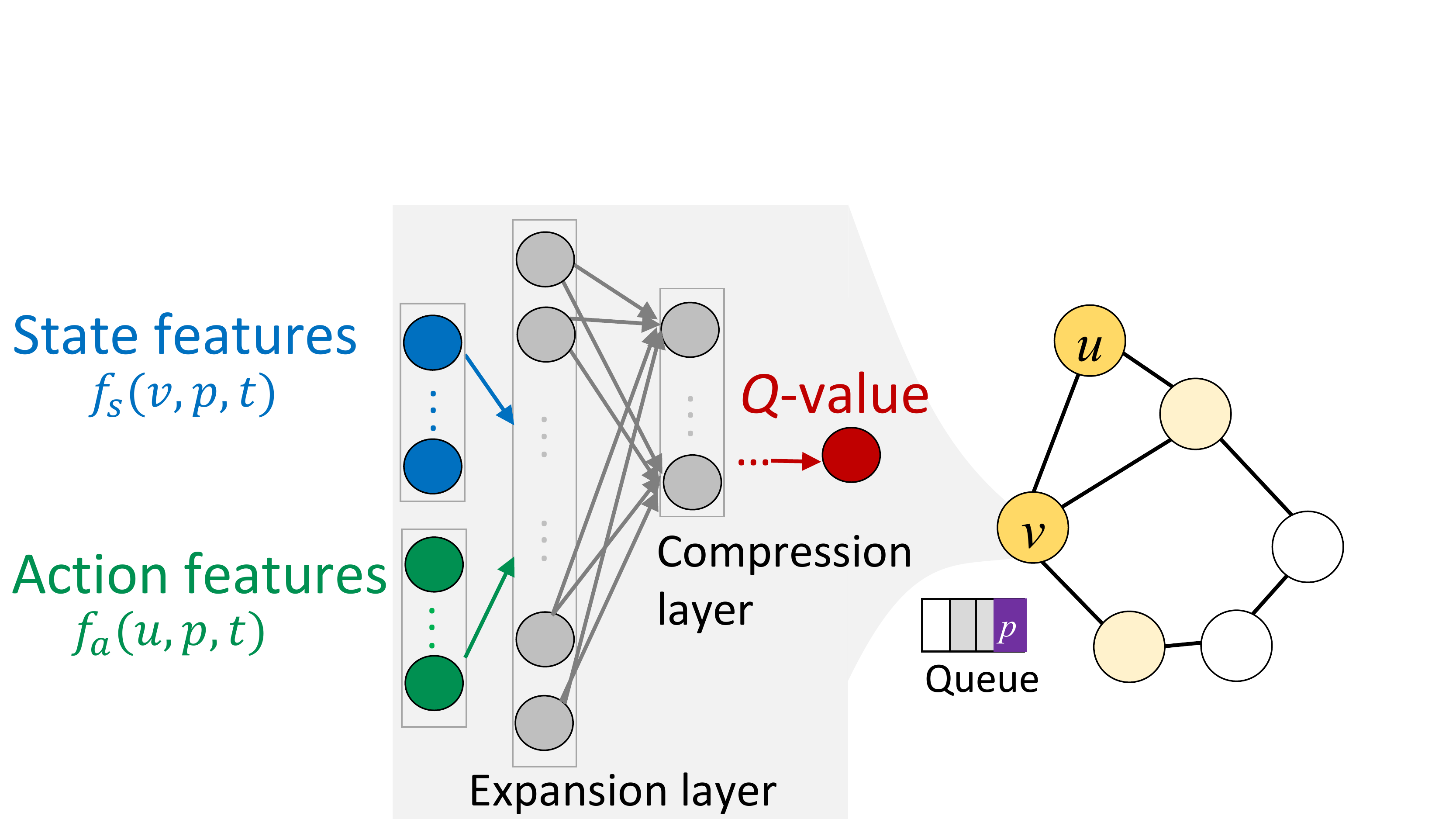}}}
   \hspace{0.1in}
   \subfigure[The Q-learning backup diagram for a packet $p$ which goes from device $w$ to $v$ at timestep $t_i$ and then must choose among neighboring devices $u \in \mathcal{A}(v)$ at timestep $t_j$.]{\hbox{\includegraphics[width=2.1in]{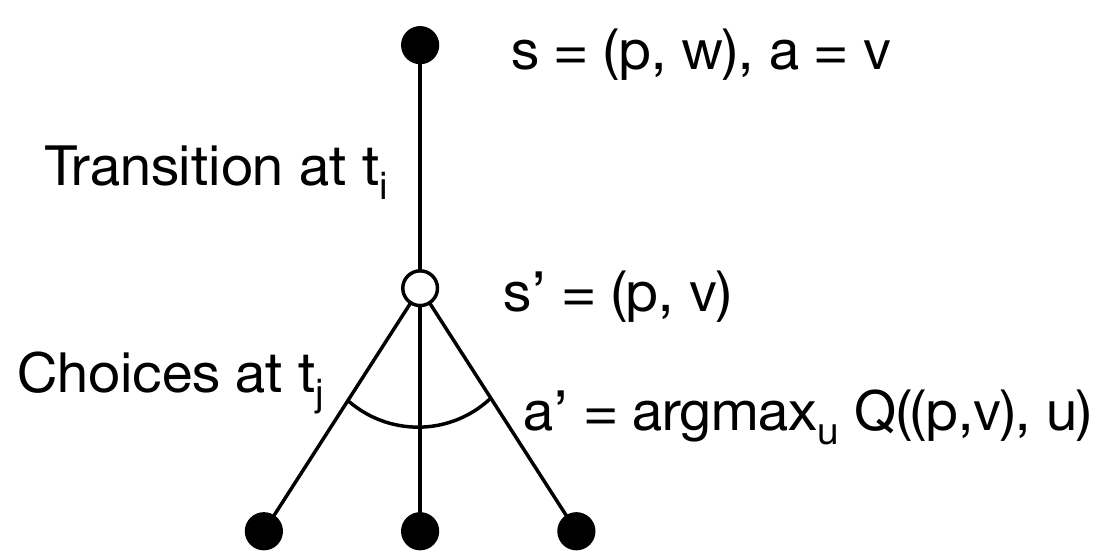}}}
    }
\vspace{-0.1in}
\caption{{\small Diagrams illustrating our algorithm operation.}}
\label{arch}
\vspace{-0.1in}
\end{figure*}
\mysection{A Reinforcement Learning Model for Routing}
\label{sec:model}

The goal of reinforcement learning (RL) \cite{sutton2018reinforcement} is to learn to choose actions to maximize expected future reward. 
RL uses a Markov decision process (MDP) to describe an agent's environment. An MDP comprises a set of states ($S$), a per state set of actions ($\mathcal{A}(s)$), a reward function, and a state transition function. State transitions are assumed to be Markovian: the probability of the next state $s' \in S$ depends only on the current state $s \in S$ and action $a \in \mathcal{A}(s)$. 
RL assumes that these state transition probabilities are {\em not} known, but that samples from the environment can be generated. 
The $Q$-value  for each $(s, a)$ pair estimates the expected future reward for an agent, when starting in state $s$ and taking action $a$. To learn, the agent observes $(s, a, r, s')$ at each time step, where $r$ is the immediate reward. The $Q$-value function is then updated via:
\vspace{-0.05in}
\begin{align*}
    Q(s,a) &\leftarrow Q(s, a) + \alpha\left[r + \gamma \cdot \max_{a' \in \mathcal{A}(s')}Q(s', a') - Q(s, a)\right] 
\end{align*}
where $0 \leq \gamma \leq 1$ is a discount factor 
which indicates the relative value of present and future rewards, and $0 < \alpha \leq 1$ is a learning rate. 
Once learned, the optimal action in state $s$ is the one with the  highest $Q$-value.

When the MDP has a small number of states and actions, an RL agent can learn a $Q$-value function 
using Q-learning (see \cite{watkins1992q}).
When the state space is too large for exact computation of the $Q$-values, function approximation may be used to find approximate $Q$-values. 
In this paper, we use deep neural networks (DNNs) \cite{bengio2009learning} for function approximation. Each state $s$ and action $a$ are translated into a set of features via the functions $f_s(s)$ and $f_a(a)$. These features are then used as input to the DNN, to produce as output an approximate $Q$-function $\hat Q(f_s(s), f_a(a))$. 
When training the DNN for each observation $(s, a, r, s')$, we use  $f_s(s),f_a(a)$ as input and $y$, defined as follows, as output (sampling from ${\hat Q}$):
\vspace{-0.05in}
\begin{align}
\label{eqn:targetonestep}
    y &= r + \gamma \cdot \max_{a' \in \mathcal{A}(s')}\hat Q(f_s(s'), f_a(a')).
\end{align}
\vspace{-0.15in}


\mysubsection{Formulating an MDP for Routing}

Consider a wireless device $v$ and packet $p$ at the front of $v$'s queue. 
State features are derived from the packet $p$, device $v$, and neighbors of $v$. Action features are derived from the  neighbor $u$ under consideration for the role of packet $p$'s next hop. When links are dynamic, only those neighbors for which there is currently a link are considered as possible actions. 
The specific  features we use are relational (see \S\ref{sec:features}).
The next hop for $p$ is found
by inputting the features into a trained DNN (see \S\ref{sec:training}). 
The output of the DNN is a Q-value 
indicating the expected future reward of choosing neighbor $u$ as packet $p$'s next hop given the state and action features.

Learning the decision model that will be applied for the packet agent at each device is computationally expensive, so training takes place offline. Once the decision model has been learned, it is copied to each device and ``frozen." During use the parameters of the model are fixed, though the feature inputs vary.
This allows fast decision making at each device. 

\mysubsection{Relational State and Action Features}
\label{sec:features}
Relational features are independent of the network topology and traffic on which the DNN is trained.  
All packets in the network  use the same relational features 
for routing decisions, though the values of those features for individual packets may differ. A simple example of such features is given in Fig. \ref{arch}(a).
We  omit features like device ID, packet destination ID,  and any other features containing identifying information specific to a device or packet that would prevent generalization.

In this work, we use a finite set of easy-to-calculate state and action features for the per-packet state $s$ and action set $\mathcal{A}(s)$, defined by the following functions. 

\smallskip
\noindent{\em{State features.}}  For a packet $p$ at device $v$ at time $t$, with one-hop neighbor set $Nbr(v)$, the state features are a function,  ${f}_s(p, v, t)$ computed from the following packet, device, and neighbor features at time $t$.

\begin{enumerate}
\item {\em Packet features,} ${f}_{pkt}(p, v, t)$. These are features derived from the packet $p$ itself. We use (i) $p$'s TTL field and (ii) $p$'s location in device $v$'s queue.

\item {\em Local device features,} ${f}_{device}(v, p, t)$. These are features derived from the device $v$ at which the packet is currently located. We use (i) the estimated distance from device $v$ to the packet's destination $dest(p)$,
(ii) $v$'s queue length,  (iii) $v$'s queue length considering only packets destined to $dest(p)$, and (iv) $v$'s  degree.

\item {\em Aggregated neighbor features,} ${f}_{nbr}(Nbr(v), p, t)$. These are features aggregated over all neighbors of device $v$. We first compute the local features for each neighbor, $f_{device}(u, p, t)$, for each $u \in Nbr(v)$. Then we compute the minimum, mean, and maximum of these features. This is similar in spirit to the aggregation function in a graph neural network \cite{google-gnn}.
\end{enumerate}

The state features then comprise the three sets of features:
\vspace{-0.075in}
\begin{eqnarray*}
{f}_s(p, v, t) &= & {f}_{pkt}(\cdot) \cup {f}_{device}(\cdot) \cup {f}_{nbr}(\cdot). 
\end{eqnarray*}
Let ${f}_i$ be the value of feature $i$ and $f_i^{max}$ be its maximum.  We use
the normalized features $f_i = ({f}_i + 1) / (f_i^{max}+1)$ as the input to the DNN, where we add 1 to both numerator and denominator so that no feature has value 0. 
In the rest of the paper, $f_s(\cdot), f_{pkt}(\cdot), f_{device}(\cdot), f_{nbr}(\cdot)$ all refer to the normalized features. We set the maximum destination distance to $N$, the maximum queue length to $B$, the maximum degree to $N$, and the maximum TTL to $L$, see Table \ref{tab:parameters}.


\smallskip
\noindent{\em{Action features.}} Each action at time $t$ selects a next hop for the packet $p$ that is at the front of device $v$'s queue. A packet can choose either to stay at its current device or transition to one of the neighboring devices.
Let $\mathcal{A}(v) = Nbr(v) \cup \{v\}$. Actions are represented via relational features that abstractly represent these choices. For packet $p$ considering moving from $v$ to $u \in \mathcal{A}(v)$, the features for action $u$ are given by  $f_{a}(u, p, t)$, which corresponds to the local device features of  $u$:
\begin{eqnarray*}
f_{a}(u, p, t) & = & f_{device}(u, p, t). 
\end{eqnarray*}
\vspace{-0.2in}

\mysubsection{Reward Function} \label{sec:reward}
RL agents  optimize a reward function by prediction of the expected future reward for each state $s$, action $a$ pair.
We divide states into three categories: 
(1) {\em delivery} states, in which a packet is delivered to its destination, (2) {\em drop} states, in which a packet is dropped, and (3) {\em transition} states, in which a packet either stays at its current device or  is transmitted to a neighbor that is not its destination.
Our reward function, $r$, is then:
\vspace{-0.05in}
\begin{eqnarray*}
~~r_{delivery} = 0, r_{transition} = -1, r_{drop} = r_{transition}/(1-\gamma) 
\end{eqnarray*}
where $\gamma \in [0,1]$, as described earlier, is the RL discount factor.
In \S\ref{sec:pkt-agent}, we describe a packet-centric view that formulates routing as an episodic task that terminates when a packet is delivered or dropped. The drop reward $r_{drop}$ is defined to be equivalent to receiving $r_{transition}$ for infinite timesteps.

\mysubsection{Packet Agents vs. Device Agents}
\label{sec:pkt-agent}

An agent's experience consists of $(s, a, r)$ tuples that are chained together into time sequences by the next state, $s^\prime$.
There are two natural ways to do this. In {\em device-centric decisions}, each device is an agent and independently makes a decision about where to forward the packet at the front of the queue. In {\em packet-centric decisions}, each packet that travels through the network is an agent and independently makes a decision when it reaches the front of a device's queue; it may choose to stay at the current device or move to a neighbor.
In both cases, the state and action features are the same. The difference is in how states and actions are chained together in time sequences: either all experiences from the same packet form a sequence, or all experiences from the same device form a sequence. 
While the device-centric approach may seem more natural  since we typically think of wireless devices as making routing decisions, here we use the packet-centric approach (see Fig. \ref{arch}(b)) as
it provides a more natural way to propagate reward, which is defined based on packet states, back to the previous time steps and actions that helped deliver the packet (or not).

Packet-centric decision-making can be viewed as a multi-agent problem, as each packet interacts with others while attempting to greedily optimize its own travel time. To reduce computational complexity,  however, we do not use a global cooperative reward function. Instead, we have each device queue enforce fairness among its packets: e.g., only the packet at the front of the queue gets to choose to move to another device at each time step. 

\mysubsection{Actions vs. Options}
Actions in Q-learning typically take only one time step to complete.
Routing actions, however, often involve multiple time steps: e.g., after a packet arrives at a new device, it waits for some time in the device's queue, with no opportunity to make a routing decision. 
This scenario is a natural case for extended-time actions, or options \cite{sutton1999between}. The time interval that the packet waits is treated as a single option, which takes a variable amount of time.
This approach requires less data to be collected, and allows Q-learning to proceed more quickly. 

The sample estimate of expected return for an  option that starts at time step $t_i$ and ends at time step $t_j$ is (see \cite{sutton1999between}): 
\vspace{-0.05in}
\begin{align*}
y &=\sum_{k=t_i}^{t_j-1} \gamma^{k-t_i} \cdot r_{k} + &\left [\gamma^{(t_j - t_i)} \cdot \max_{a' \in {\mathcal{A}(s_{t_j})}} Q(s_{t_j}, a') \right ]
\end{align*}
where $r_k$ is the reward at time step $k$, and $s_{t_j}$ is the state encountered at time $t_j$, with $\mathcal{A}(s_{t_j})$ its actions. We use this as the output target $y$ for the neural network, replacing Eq. (\ref{eqn:targetonestep}). 

Here, we consider only reward functions that are constant for every timestep over the life of the option (see \S\ref{sec:reward}), so that all $r_{k} = r_{c}$, where $r_c$ is one of our three reward types. 
The sample return for an option starting at time $t_i$ and ending at time $t_j$ with constant per-time step reward $r_c$ is then:
\vspace{-0.05in}
\begin{align}
\label{eqn:targetoptions}
y &= R(t_j -t_i, r_c) + \gamma^{(t_j-t_i)} \cdot \max_{a' \in {\mathcal A}(s_{t_j})} Q(s_{t_j}, a')
\end{align}
\vspace{-0.15in}
where
\vspace{-0.1in}
\begin{align*}
    R(t_j -t_i, r_c) & = r_c \cdot \frac{1-\gamma^{(t_j-t_i)}}{1-\gamma}.
\end{align*}
There are two types of option in this domain: terminal (packet delivery or drop) and non-terminal (transitions from one device to another). On packet delivery or drop, the option takes only a single time step and the next state $s_{t_j}$ is the terminal state. The sample of return for delivery is: 
\vspace{-0.05in}
\begin{align*}
y &= R(t_j - t_i, r_{delivery}) + \gamma^{(t_j - t_i)} \cdot \max_{a' \in {\mathcal A}(s_{t_j})} Q(s_{t_j}, a') \\
&= r_{delivery} \cdot \frac{1-\gamma^{1}}{1-\gamma} +0 \\
&= r_{delivery},
\end{align*}
and $r_{drop}$ transitions similarly reduce to the immediate reward.

\noindent{On non-terminal transitions the sample of return is:}
\vspace{-0.05in}
\begin{align*}
y &= R(t_j -t_i, r_{transition}) +
\gamma^{(t_j-t_i)} \cdot \max_{a' \in {\mathcal A}(s_{t_j})} Q(s_{t_j}, a').
\end{align*}
\vspace{-0.15in}

Every packet that remains in a device queue at the end of a training  round has an unfinished option. 
We remove such options from the data. 
As more data accumulates, including the end of the option, the newly finished options are used.

From this point on, to be consistent with \S\ref{sec:features}, we use ``actions" rather than ``options" to refer to extended-time actions.

\mysubsection{Function approximation}
\label{sec:training}

We use a deep neural network (DNN) \cite{bengio2009learning} to approximate the value function. 
Since we assume all devices use the same DNN to make decisions, data from all devices can be pooled  into a single large training set.  During testing, each device independently uses its own copy of the DNN for testing.

Our DNN architecture has 4 layers as in Fig. \ref{arch}(b): input, expansion, compression, and output. 
Let $F=|f_s(\cdot)| + |f_a(\cdot)|$ be the number of input features and thus the size of the input layer. The expansion layer has $10F$ neurons and the compression layer has $F/2$ neurons.
The input to our DNN is a state, action pair (s, a) represented through their feature vectors.  The output is a single neuron which estimates the Q-value of (s, a).
For packet $p$ at device $v$ considering moving to device $u$ at time $t$, this takes the form of a Q-function with feature inputs:
$Q(f_s(p, v, t), f_a(u, p, t)).$
We use the following shortened notation:
\vspace{-0.1in}
\begin{eqnarray*}
Q((p,v),u, t) &=&Q(f_s(p, v, t), f_a(u, p, t)).
\end{eqnarray*}
\vspace{-0.2in}
\begin{table}[t]
\centering
\caption{{\small Information in  row of data for (state, action) pair.}}\smallskip
\vspace{-0.1in}
\begin{tabular}{ll}
Symbol             & Meaning \\
\hline \hline
$id(p)$ & id of packet $p$ \\
$id(v)$ & id of device $v$\\
$t_i$ & time that packet arrives at device $v$\\
$t_j$ & time that packet departs device $v$ \\
$f_s(p, v, t_j)$ & state features at $t_j$ for device $v$\\
$f_a(u, p, t_j)$ & action features at $t_j$ for neighbor $u$\\
$r(p, v)$ & observed reward  \\
$b$  & flag: whether $u$ was selected at $t_j$\\
\hline
Calculated columns\\
$Q((p, v), u, t_j)$ & estimated value of going from $v$ to $u$\\
$\max_{u \in {\mathcal A}(v)} Q((p, v), u, t_j)$ & max value over all $u$ rows for $(p, v)$ \\
\hline
Joined columns \\
$f_s(p, w, t_i)$ & state features at $t_i$ for packet $p$ device $w$\\
$f_a(v, p, t_i)$ & action features at $t_i$ for neighbor $v$\\
\hline
\end{tabular}
\label{tab:data-row}
\vspace{-0.15in}
\end{table}

{\em Action selection.} Different numbers of actions are available to packets at different devices. 
Thus, when packet $p$ makes a decision at device $v$, the feature set $(f_s(\cdot), f_a(\cdot))$ for each device $u \in \mathcal{A}(v)$ is fed into the DNN as input to obtain a list of estimated $Q((p, v),u)$ values, one Q-value for each possible action. These Q-values can then be fed into any action selection mechanism. Here, we use $\epsilon$-greedy with $\epsilon$ set as in Table \ref{tab:parameters}.

{\em Data collection.} In each training round, we gather a sequence of experience tuples $\cdots, (s, a, r), (s', a', r'), \cdots$, using the current DNN to choose actions. 
We record data only for
packets that make  a decision, i.e., the packet $p$ at the front of the queue at each device $v$. 
For every possible action $u \in \mathcal{A}(v)$ available to $p$ at $v$ at time step $t_j$, 
we record the information  in Table \ref{tab:data-row} as a row in our data.
The row that contains the action selected by $p$ has flag $b$ set to one. 
The columns marked Calculated and Joined are added later 
by the algorithm.

{\em Training.} For each packet $p$ which transitions from device $w$ to device $v$, arriving at $v$ at time step $t_i$ and departing at $t_j$, we use Algorithm \ref{alg:training} to do the ``backup" shown in Fig. \ref{arch}(c). This update improves the estimate of $Q((p, w), v)$.
Non-relational columns are used to find the device $w$ that packet $p$ departed at time $t_i$ (to chain together $(s, a, r, s')$). However, only $f_s(p, v, t_j)$ and $f_a(u, p, t_j)$ are used as input to the DNN to estimate $Q((p, v), u, t_j)$ for each $u \in \mathcal{A}(v)$ when calculating the target values. We use 
Eq. (\ref{eqn:targetoptions})
to calculate the training target. 

\newlength{\textfloatsepsave} \setlength{\textfloatsepsave}{\textfloatsep} \setlength{\textfloatsep}{0pt} 
\begin{algorithm}[t]
\begin{footnotesize}
\caption{{\small Steps executed for one round of training. 
}}
\label{alg:training}
\SetKwInput{kwInput}{Input}
\SetKwBlock{kwInit}{Initialization}{end}
\SetKwBlock{kwMain}{Forward Inclusion}{end}
\SetKwInput{kwOutput}{Output}
\SetAlgoLined
\kwInput{
\par
$nn$: a new randomly initialized DNN for this round
\par
$data$: collected from time 0 up to and including this round} \par
\par
\For{k = 1 to \# of Q-learning iterations}{
\underline{\bf Calculate target $y$} \par
    Estimate $Q((p, v), u, t_j)$ for all rows using $nn$ and add column to $data$\par
    Take max over actions: $\max_{u \in {\mathcal A}(v)} Q((p, v), u, t_j)$, add column \par
    Add $y = R(t_j - t_i, r(p,v)) + \gamma^{t_j-t_i} \max_{u \in {\mathcal A}(v)} Q((p, v), u, t_j)$ \par
\underline{\bf Find previous device $w$ and filter} \par
    Filter to select those rows with $b = 1$ (only actions chosen) \par
    Join $data$ with itself on $id(p)$ and $t_i$ matched with $t_j$, labelling \\ the $t_i$ device $w$, and $t_j$ device $v$ 
    \par
    Remove rows with unfinished actions \par 
\underline{\bf Fit $nn$} \par
    Improve estimate of $Q((p,w), v, t_i)$ using: \\ input $f_s(p, w, t_i), f_a(v, p, t_i)$, target $y$ 
}
\end{footnotesize}
\end{algorithm}
\setlength{\textfloatsep}{\textfloatsepsave}

\mysection{Performance Evaluation}
\label{sec:evaluation}

We evaluate our approach 
using a 
discrete-time packet-level network simulator that we have implemented in Python3. This simulator provides the environment in which the DRL agents are trained and all routing algorithms are tested. Tables \ref{tab:parameters} to \ref{tab:traffic} show our simulation parameters.
We use Keras v.2.3.1 \cite{chollet2015keras} and Tensorflow v.1.14.0 \cite{tensorflow2015-whitepaper} to implement the  DNN. Training and testing was done using the MIT SuperCloud and Lincoln Laboratory Supercomputing Center \cite{reuther2018interactive}.

\mysubsection{Simulation Settings}
\label{sec:sim-settings}

Our goal is to identify the wireless scenarios for which our DRL approach performs well (i.e., delivers the most packets with low delay)
and to test how well a DRL agent trained on one scenario is able to generalize its learned routing policy to  unseen scenarios. We thus explore a wide range of scenarios that differ in topology, link dynamics, and  traffic.

{\em Topologies.} As shown in Fig. \ref{fig-results-policies},  we consider two types of network topologies: 
(i) a square grid lattice, and 
(ii) a geometric random graph where devices are randomly placed in a unit square, and two devices are connected by a link if they are within a given
transmission radius. 

{\em Link dynamics.} To model link dynamics, we use a 2-state Markov model. We assume links are i.i.d and stay up from one time step to the next with probability $\alpha$ (and transition down with probability $1-\alpha$), and stay down from one timestep to the next with probability $\beta$ (and transition up with probability $1-\beta$).
For a given topology, we initialize the up and down states of links based on the steady-state link probability for this 2-state model,  $\pi = (1-\alpha)/(2-\alpha-\beta)$.

{\em Medium access control.} 
On each timestep, we  loop through all devices in random order and allow each  device to transmit a single packet that was received or generated in a previous timestep. Doing this imposes per-device capacity constraints. 

\begin{table}[t]
\centering
\caption{{\small Simulation parameters}}\smallskip
\vspace{-0.12in} 
\begin{tabular}{lll}
Symbol             & Meaning                    & Value \\
\hline \hline
$N$                & \# of network devices      & 9 to 100 \\
$\epsilon_{train}$ & Training exploration rate  & .1 \\
$\epsilon_{test}$  & Testing exploration rate   &  0 \\
$\gamma$           & RL discount rate           & 0.99 \\
$r_{transition}$   & Transition reward          & -1 \\
$r_{drop}$         & Drop reward                & $r_{transition} / (1-\gamma)$ \\
$r_{delivery}$     & Delivery reward            & 0 \\
$L$                & Packet time-to-live        & 200 \\
$B$                & Maximum queue size         & 50 or $50N$ \\
$T_{train}$        & \# of training timesteps   & 30,000 or 49,000 \\
$T_{test}$         & \# of testing timesteps    & 100,000 \\
$T_{round}$        & \# of timesteps per round  & 1000 \\
\hline
\end{tabular}
\label{tab:parameters}
\vspace{-0.1in}
\end{table}

{\em Traffic generation.} Sources and destinations are selected uniformly randomly, with  the constraint that a source never equals its associated destination. 
The flows present then change over time: new flow arrivals are generated according to a Poisson distribution with parameter $\lambda_F$; flow durations are generated by sampling an exponential distribution with parameter $\lambda_D$. 
Packet arrivals are generated according to a Poisson distribution with parameter $\lambda_P$, where $\lambda_P$ is the average number of new packets generated per timestep on a given flow.
A simulation starts with $\lambda_F \lambda_D$ initial flows.

{\em Queue size.} Each device in the network maintains a packet queue with a maximum buffer size, $B$, beyond which additional packets are dropped.

\subsection{Training and Testing Scenarios}  

We organize the wireless network scenarios that we consider along  three dimensions, connectivity,  size, and and congestion.

\subsubsection{Network connectivity}

As in \cite{manfredi2011understanding}, network connectivity influences the kind of routing strategy that is appropriate. Ad hoc routing strategies work well  in networks that are mostly well-connected, while delay tolerant routing strategies work well in networks that are mostly disconnected. 
To measure network connectivity
we use {\em algebraic connectivity}, defined as the second-smallest eigenvalue of the normalized Laplacian matrix of a graph \cite{chung1996lectures}. 
The larger the value, the more well-connected is the topology.  
When algebraic connectivity equals 0, the network is disconnected. 
Here, we vary network connectivity (and thus algebraic connectivity) in terms of  (i) {\em link dynamics},  and (ii) {\em path redundancy}.

To vary {\em link dynamics} 
we vary the values of $\alpha$ and $\beta$
for our 2-state link model in \S\ref{sec:sim-settings}.
For  large $\alpha$ and  small $\beta$, the network is connected; as $\alpha$ decreases and $\beta$ increases, the probability that a contemporaneous end-end path exists between two devices decreases. 
The different link dynamics we  use are shown in Table \ref{tab:dynamics}. The special case with $\alpha=1$ and $\beta=0$ has no link dynamics, and is referred to as {\em static}. For certain settings of $\alpha$ and $\beta$ (e.g., when $\alpha=0.5$ and $\beta=0.4$), the network is mostly disconnected and is referred to as  {\em delay tolerant}. When there are some link dynamics, but the network is nonetheless mostly connected, 
is referred to as {\em dynamic}.

To vary  {\em path redundancy} we vary the topology.
Topologies with high redundancy should have shorter paths, and  better handle  congestion.
The lattice topology is relatively sparsely connected, while the random  topology is densely connected. As in Table \ref{tab:dynamics}, we consider two random topologies, one with a transmission radius of 0.5, and one with a radius of 0.3.

\begin{table}[t]
\centering
\caption{{\small Network topologies and link dynamics}}
\label{tab:dynamics}
\vspace{-0.05in}
\begin{tabular}{lllll}
Network scenario                         & radius     & $\alpha$  & $\beta$      & $\pi$ \\
\hline \hline
Static lattice topology                  & -     & 1.0       & $0.0$    & 1.0 \\
Dynamic lattice topology                 & -     & 0.8       & $0.2$    & 0.8 \\
Delay tolerant lattice topology          & -     & 0.5       & $0.4$    & 0.55   \\
Static random geometric topology         & 0.5   & 1.0       & $0.0$    & 1.0 \\
Delay tolerant random geometric topology & 0.3   & 0.5       & $0.4$    & 0.55   \\
\hline
\end{tabular}
\end{table}

\begin{table}[t]
\centering
\vspace{-0.1in}
\caption{{\small Network traffic scenarios.}}
\vspace{-0.05in}
\begin{tabular}{llll}
Traffic scenario  &  $\lambda_F$      & $\lambda_D$ & $\lambda_P$ \\
\hline \hline
$Low$ traffic congestion        & $.002 N / 25$    & 5000        & .05  \\
$High$ traffic congestion       & $.002 N / 25$    & 5000        & .2  \\
\hline
\end{tabular}
\label{tab:traffic}
\vspace{-0.1in}
\end{table}

\subsubsection{Network size}
Varying the network size, $N$, affects connectivity differently depending on the underlying topology. In our testing results, we vary $N$ from 9 to 100; we use two sizes, $N=$ 25 and $N=64$, for training. 
The testing results below are using the $N=64$ training results as the $N=25$  training results do not generalize as well to larger networks.
For the lattice, increasing the network size decreases connectivity. For the random geometric topology, since devices are always distributed within the unit square, increasing the network size increases connectivity.

\begin{figure}[t]    
\centerline{\hbox{
    \subfigure[\% of Packets delivered]{\includegraphics[width=1.7in]{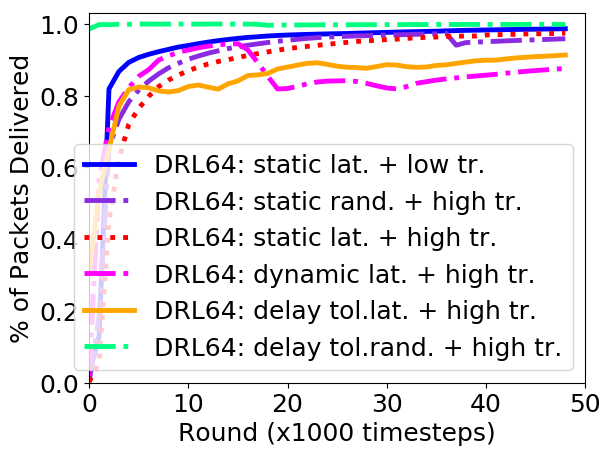}}
     \subfigure[Total packets generated]{\includegraphics[width=1.78in]{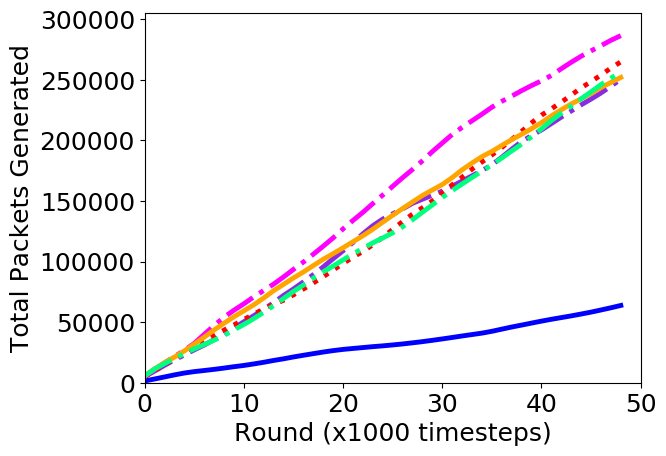}}}}
\vspace{-0.05in}
\centerline{\hbox{
    \subfigure[Delay per packet]{\includegraphics[width=1.72in]{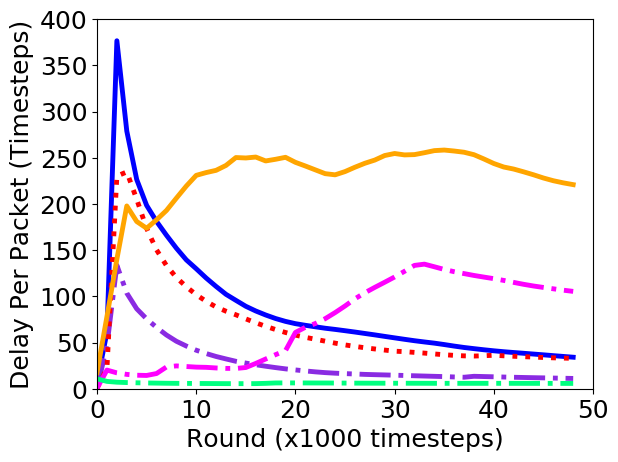}}
    \subfigure[Average queue length]{\includegraphics[width=1.65in]{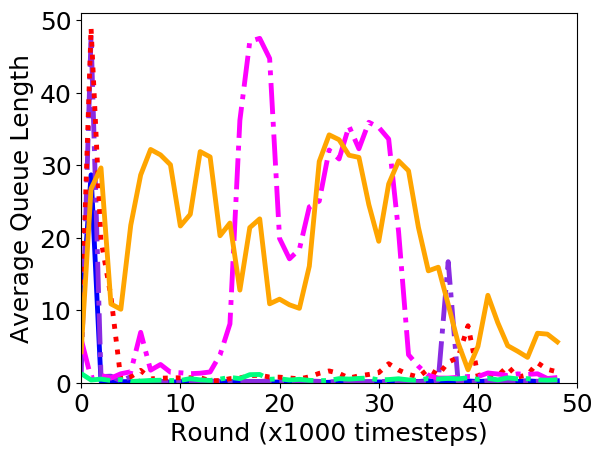}}
}}
\vspace{-0.05in}
\caption{{\small Training performance of DRL agents, for $N=64$. The legend indicates the training conditions.
}}
\label{fig-results-training}
\vspace{-0.2in}
\end{figure}

\subsubsection{Network congestion} 
We consider two traffic scenarios, $low$ and $high$, shown in Table \ref{tab:traffic}.
For both scenarios, the amount of traffic generated varies over time due to the Poisson distributed arrivals of flows and packets. During periods of increased traffic, there is correspondingly increased congestion.

When these traffic scenarios are used with other  topologies, link dynamics,
and values of $N$, the amount of traffic congestion they generate will vary.
For instance, as network size increases (i.e., $N$), there may be relatively more congestion: e.g.,  congestion scales super-linearly in a lattice topology with uniform random traffic. 
Similarly, for the same traffic scenario on the same topology, the introduction of link dynamics will decrease the available bandwidth and increase congestion.

\subsubsection{Training and testing scenarios} 
\label{sec:scenarios} 
 We vary connectivity (which results in varying the destination distance and neighbor features, see Fig. \ref{fig-results-algebraic-connectivity}(a)) and congestion level (which results in varying  the queue length feature, see Figs. \ref{fig-results-algebraic-connectivity}(b) to (d)). 
We thus obtain the  following six scenarios, which use the parameter settings shown in Tables \ref{tab:dynamics} and \ref{tab:traffic}:
(i) {\em static lattice + low traffic}, 
(ii) {\em static lattice + high traffic}, 
(iii) {\em dynamic lattice + high traffic},
(iv) {\em delay tolerant lattice + high traffic}, 
(v) {\em  static random + high traffic}, and
(vi) {\em delay tolerant random + high traffic}.
We consider only one low traffic scenario since the high traffic scenarios are more challenging.

\subsection{Routing Algorithms}  
We compare the performance of the following three routing algorithms.
All  algorithms use only local information (such as distance or destination queue length) obtained from a device's neighbors during testing. 
We do not model control packets since local (1-hop) traffic typically causes relatively little congestion compared to non-local ($>1$-hop) traffic \cite{santivanez2001making, ramanathan2010scalability}.

\subsubsection{Shortest path routing (SP)}
We implement shortest path routing as a distance vector algorithm using hop count as cost.  
We modify the algorithm slightly to accommodate link dynamics.
Because devices are stationary though the links present may change, we assume that once a device has a link to a  neighbor device, that link continues to be present in the distance calculations. Because link changes are i.i.d., this means the distance vector algorithm we use converges on the true shortest path distance in terms of hop count. However, only those neighbors for which there are links present are considered as possible next hops when a routing decision is made. For each packet, then, the next hop for the shortest path that is currently available is chosen. If no next hop is available, then the packet stays at the device.

\subsubsection{Deep reinforcement learning (DRL)}
To train, our algorithm can use either simulated network data (as we do here) or historical data recorded from a network of interest. 
A DRL agent estimates 
the distance feature using the same distributed distance vector algorithm that is used by shortest path routing. 
Because packets may take very long paths while the DRL agent is learning a good policy during training, we use the relatively high TTL value of $L=200$ compared to the expected path length, to prevent packets from always being dropped before the DRL agent has had sufficient time to learn.

\begin{figure}[t]    
\centerline{\hbox{
     \subfigure[Algebraic connectivity of scenarios. Algebraic connectivity is a topology measure, so is independent of traffic.
     ]{\includegraphics[width=1.72in]{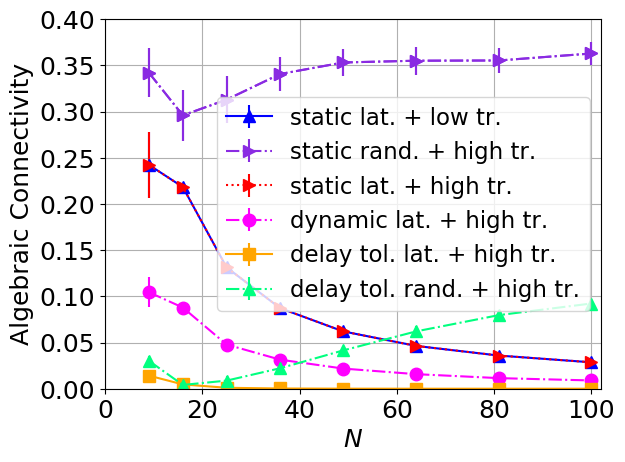}}
    \hspace{0.03in}
    \subfigure[BP queue lengths  during testing.
    BP uses  $B=50N$, so sees larger queue lengths than the other algorithms  which use $B=50$.
    ]{\includegraphics[width=1.72in]{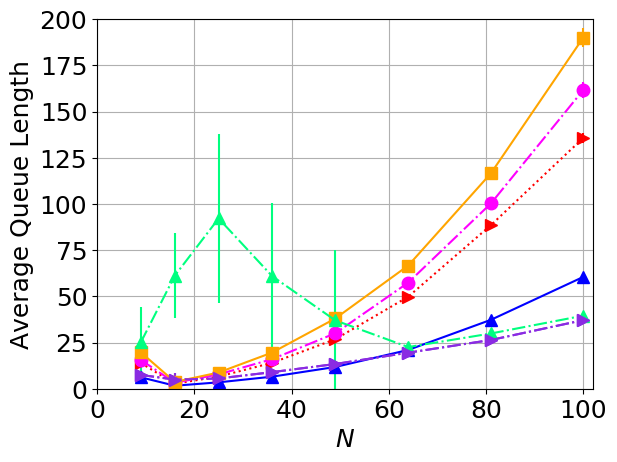}
}}}
\vspace{-0.03in}
\centerline{\hbox{
    \subfigure[SP queue lengths  during testing.]{\includegraphics[width=1.72in]{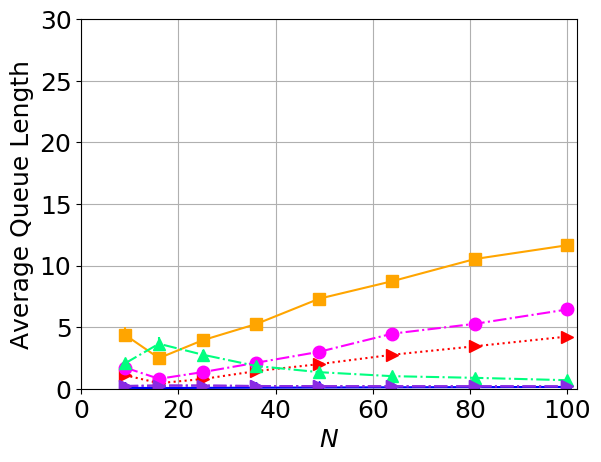}}
    \hspace{0.03in}
    \subfigure[
    Queue lengths seen for a DRL64 agent trained on the delay tol. lat. + high traffic and then tested on all scenarios.]{\includegraphics[width=1.72in]{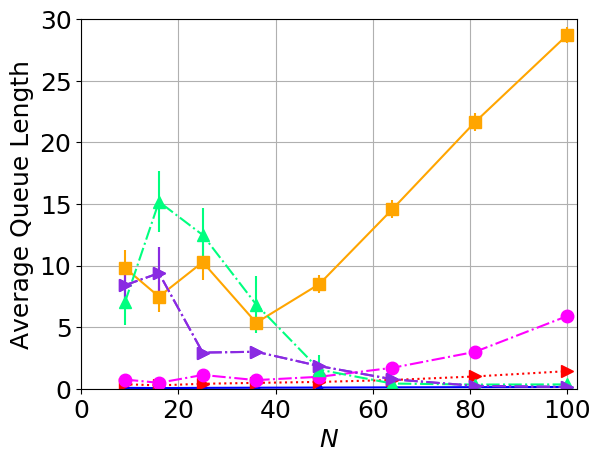}    
   
}}}
\vspace{-0.1in}
\caption{{\small Example network connectivity and congestion levels.
}}
\label{fig-results-algebraic-connectivity}
\vspace{-0.2in}
\end{figure}

\begin{figure*}[hbtp]
\centerline{\hbox{
    \subfigure[Static random + high traffic ]{\includegraphics[width=2.1in]{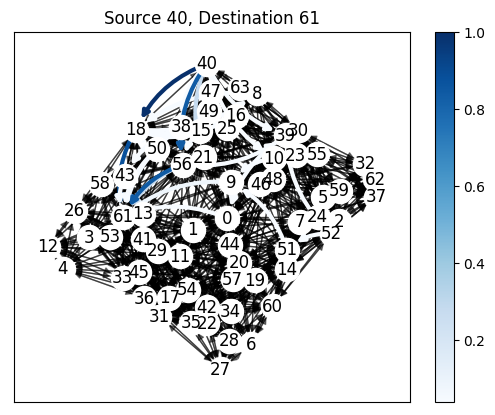}}
    \hspace{0.05in}
    \subfigure[Static lattice + high traffic]{\includegraphics[width=2.1in]{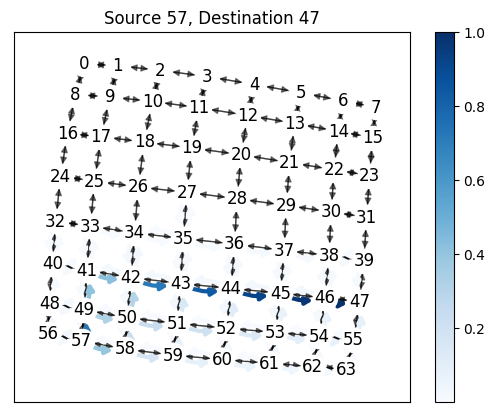}}
    \hspace{0.05in}
    \subfigure[Delay tolerant lattice + high traffic]{\includegraphics[width=2.1in]{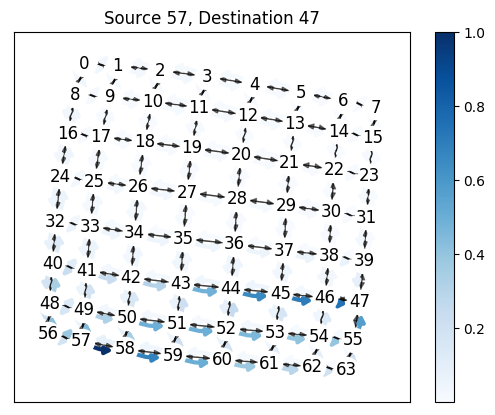}}
}}
\vspace{-0.1in}
\caption{{\small Example learned policies of DRL64 agents from training data. Plots show the number of times each action (link) is selected for a flow, normalized by the maximum number of times any link is selected for the flow. 
}}
\label{fig-results-policies}
\vspace{-0.2in}
\end{figure*}

\subsubsection{Backpressure routing (BP)}
Consider an arbitrary device $v$.  
Let $b_{d}^v$ be the number of packets destined to device $d$ in the queue at device $v$.  
For every destination $d$ of a packet in $v$'s queue, $v$ computes $b_{d}^v - b_{d}^{u}$ for the neighbors $u \in Nbr(v)$ currently available. Then $v$ finds the optimal destination $d^*$ and corresponding neighbor $u^*$, such that $b_{d^*}^v - b_{d^*}^{u^*}$ is the largest among all destinations (breaking ties arbitrarily), i.e., 
BP routes packets in the direction that maximizes the {\em differential backlog} between neighboring devices. If $b_{d^*}^v - b_{d^*}^{u^*} > 0$, then $v$ sends a packet with destination $d^*$ to $u^*$; otherwise $v$ does not send any packet. 
BP thus relies on network congestion 
to route well.

SP or our DRL approach 
 which forward the packet at the front of a device's queue,
BP chooses the best packet from anywhere in the queue to forward. Thus, BP requires the use of large queues to ensure packets are never dropped 
due to a  queue being full.
We set the maximum queue size for BP to be $B=50N$, which allows each device to (virtually) keep a separate queue for every destination, compared to $B=50$ for the other algorithms. 
We further evaluate BP with $B=50$ and find that it delivers many fewer packets than the other algorithms (results omitted in the interest of space).

\mysubsection{Results} 
\label{sec:results}
In this section, we first overview our DRL agent training performance in \S\ref{sec:training-results}. Then we evaluate how well the trained DRL agents can generalize their learned policies by testing their performance on the lattice scenarios, in \S\ref{sec:lattice}, and the random scenarios, in \S\ref{sec:random}.

In our simulation results, we plot the following  metrics.
Let $D_t$ ($G_t$) be the total number of packets delivered (generated) by round $t$, 
and let $Y_t$ be the total delay of packets delivered by round $t$. Then we compute (i) the {\em \% of packets delivered} by round $t$ with $D_t / G_t$;  (ii) the {\em delay per packet} by round $t$ with $Y_t / D_t$;
(iii) {\em average queue length} at round $t$  by averaging over all device queue lengths on the last timestep in the round; and
(iv) {\em algebraic connectivity} at round $t$  on the network topology on the last timestep in the round. 

\subsubsection{Learning curves and learned policies} 
\label{sec:training-results}
We train DRL agents for each scenario in \S\ref{sec:scenarios} for $N=64$, labeled DRL64. 
Training is divided into $T_{train}/T_{round}$ rounds, see Table \ref{tab:parameters}; we use 10 epochs and a batch size of 32.

Fig. \ref{fig-results-training}(a)  shows that when training on the  static scenarios, the DRL64 agents quickly learn policies that deliver all packets. 
In comparison, the DRL64 agents trained on the dynamic and delay tolerant lattice scenarios show 
fluctuation in packets delivered as
the number of flows varies over time. While the mean number of flows is given by $\lambda_F$, the actual number of flows at any given time can be higher (or lower) and cause queues to build up. 
The DRL64 agent trained on the delay tolerant random scenario converges quickly, however, due to the high connectivity (giving many possible paths) combined with the high traffic (increasing congestion and providing feedback on which paths not to use).
Because of the extra time needed for the DRL64 agents to converge when trained on the dynamic and delay tolerant lattice scenarios, ,  
for testing we train these DRL64 agents  with $T_{train} = 49,000$ timesteps; 
all other DRL64 agents are trained with $T_{train} = 30,000$ timesteps.

\begin{figure*}[t!]
 \centerline{\hbox{
    \subfigure[Static lattice + low traffic]{   
        \includegraphics[width=1.85in]{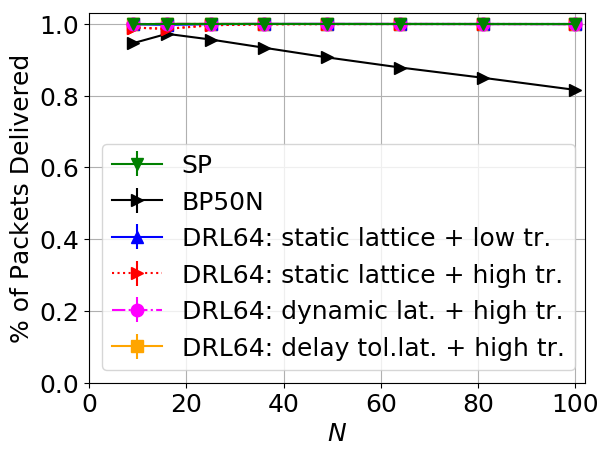}}
    \hspace{-0.19in}        
     \subfigure[Static lattice + high traffic]{   
        \includegraphics[width=1.85in]{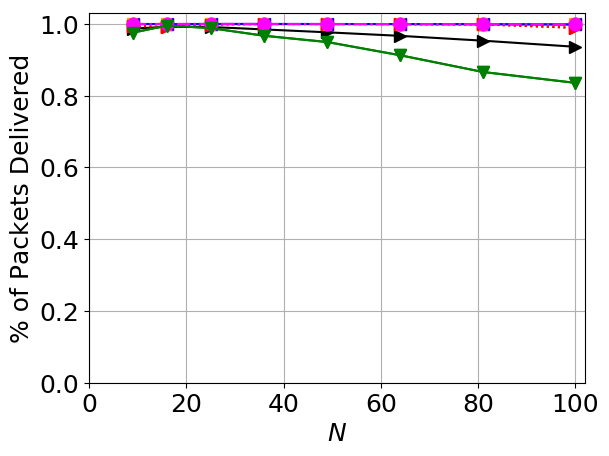}}
    \hspace{-0.19in}        
    \subfigure[Dynamic lattice + high traffic]{   
        \includegraphics[width=1.85in]{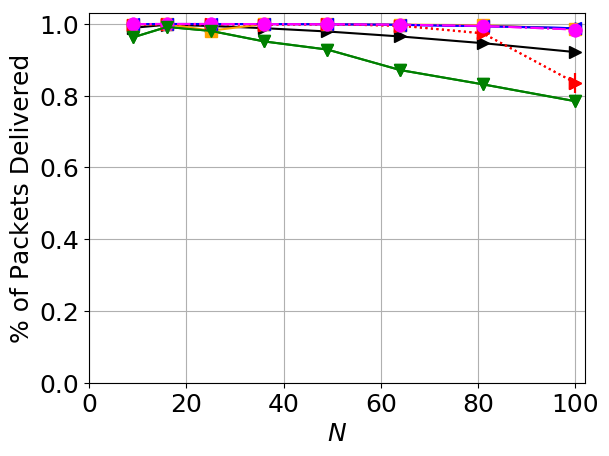}}
    \hspace{-0.19in}        
    \subfigure[Delay tolerant lattice + high traffic]{   
        \includegraphics[width=1.85in]{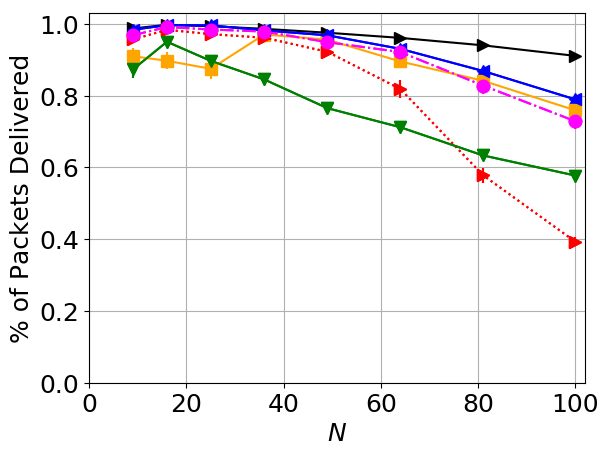}
}}}
\vspace{-0.05in}
 \centerline{\hbox{
    \subfigure[Static lattice + low traffic]{   
        \includegraphics[width=1.85in]{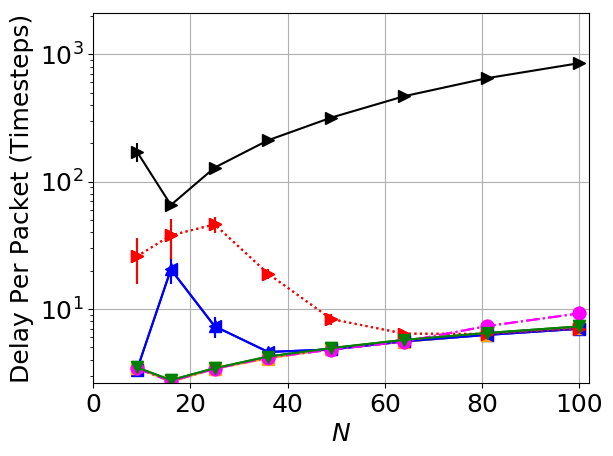}}
    \hspace{-0.19in}        
     \subfigure[Static lattice + high traffic]{   
        \includegraphics[width=1.85in]{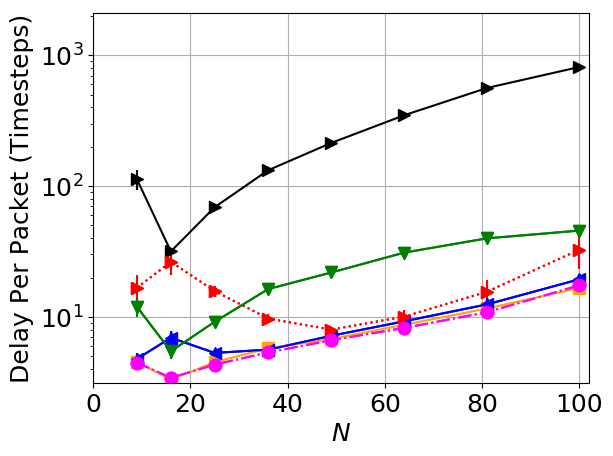}}
    \hspace{-0.19in}        
    \subfigure[Dynamic lattice + high traffic]{   
        \includegraphics[width=1.85in]{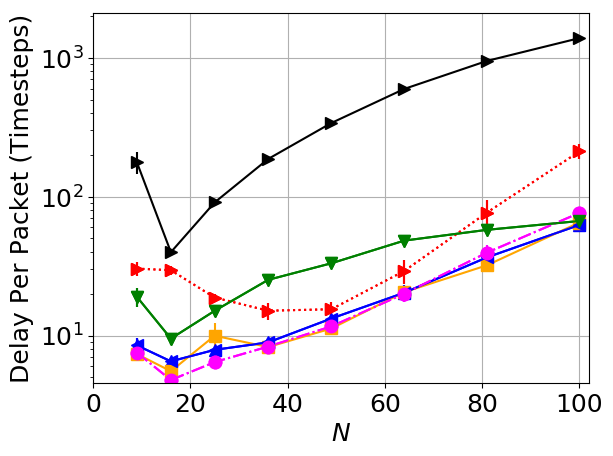}}
    \hspace{-0.19in}        
    \subfigure[Delay tolerant lattice + high traffic]{   
        \includegraphics[width=1.85in]{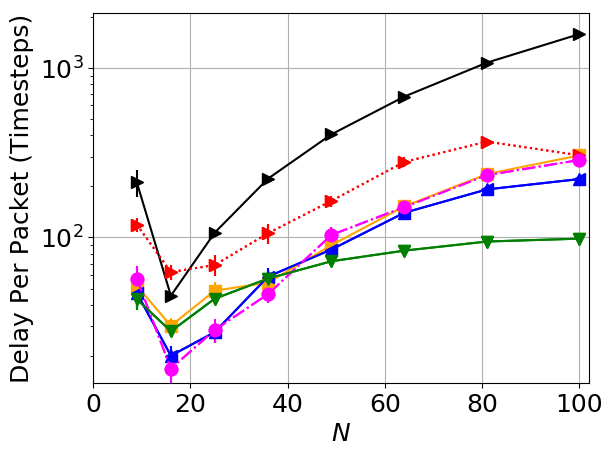}
}}}
\vspace{-0.05in}
\caption{{\small Testing generalization of DRL64 agents on the lattice topologies; connectivity decreases as $N$ increases. The training conditions (using $N=64$) are shown in the legend and the testing conditions are shown in the figure captions. 
}}
\label{fig-results-traffic-cross}
\vspace{-0.2in}
\end{figure*}

Fig. \ref{fig-results-policies} shows example learned policies. 
In the  delay tolerant lattice + high traffic scenario in Fig. \ref{fig-results-policies}(c),   the DRL64 agent  learns to distribute traffic over additional paths to the destination, which  has similarities with the  row-first column-next policy for optimal shortest path routing in a lattice \cite{barrenetxea2006lattice}.

\subsubsection{Lattice generalization} 
\label{sec:lattice}
This section shows results for when DRL64 agent training and testing are done on a lattice scenario, see Fig. \ref{fig-results-traffic-cross}.
The top row of Fig. \ref{fig-results-traffic-cross} shows the packet delivery rate while the bottom row shows packet delay for the four lattice testing scenarios (as marked in the  captions), each using all four lattice training scenarios (as marked in the legend). Each  point in
Fig. \ref{fig-results-traffic-cross} (and Figs. \ref{fig-results-algebraic-connectivity} and \ref{fig-results-traffic-cross-rand})
shows  the  95\% confidence interval computed over 50 simulation runs.

Figs. \ref{fig-results-traffic-cross}(a) and (e) show testing on the {\em static lattice + low traffic} scenario.
Due to low traffic, queues are mostly empty (see Fig. \ref{fig-results-algebraic-connectivity}), and SP is optimal.
Indeed, both the DRL64 agents and SP  deliver all packets. BP, however, delivers significantly fewer packets as network size increases, because there is insufficient traffic  for it to effectively build a congestion gradient. 

Figs. \ref{fig-results-traffic-cross}(b) and (f) show testing on the {\em static lattice + high traffic} scenario. Despite the increased traffic, the DRL64 agents are still able to deliver all packets. SP, however,  delivers significantly fewer packets.
Conversely, BP is now able to build an effective congestion gradient and  delivers more packets than SP 
but with significantly packet delay  than the DRL64 agents.

Figs. \ref{fig-results-traffic-cross}(c) and (g) show testing on the {\em dynamic lattice + high traffic} scenario.
 Although the topology and traffic is the same as in Figs. \ref{fig-results-traffic-cross}(b) and (f), due to the addition of link dynamics, the available bandwidth is reduced. Now, all DRL64 agents deliver all packets except that trained on the static lattice + high traffic. Neither SP nor BP are able to deliver all packets once the network size is sufficiently large.

Finally, Figs. \ref{fig-results-traffic-cross}(d) and (h) show testing on the {\em delay tolerant lattice + high traffic} scenario. 
This is the only lattice scenario for which the network is predominantly disconnected (see 
Fig. \ref{fig-results-algebraic-connectivity}(a)).
Due to the increased link dynamics, as $N$ increases the network becomes sufficiently congested that not all packets can be delivered.
BP now delivers the most packets in the larger network sizes because it can choose any packet in the queue to send rather than only the one at the front.  
The DRL64 agents, except that trained on the static lattice + high traffic, deliver close to 80\% of traffic for $N=100$, despite being restricted to choosing the packet at the front of the queue to send, and using the much smaller queue size of $B=50$.

\subsubsection{Random geometric generalization} 
\label{sec:random}

\begin{figure}[t]
\centerline{\hbox{
        \subfigure[Static random + high traffic]{
        \includegraphics[width=1.78in]{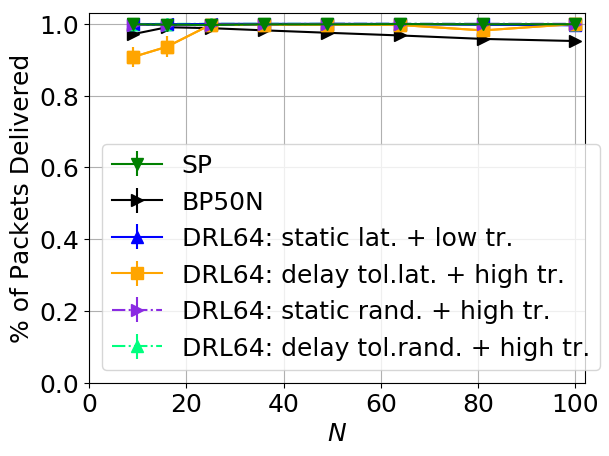}}
        \hspace{-0.18in}
      \subfigure[Delay tol. random + high traffic]{
        \includegraphics[width=1.78in]{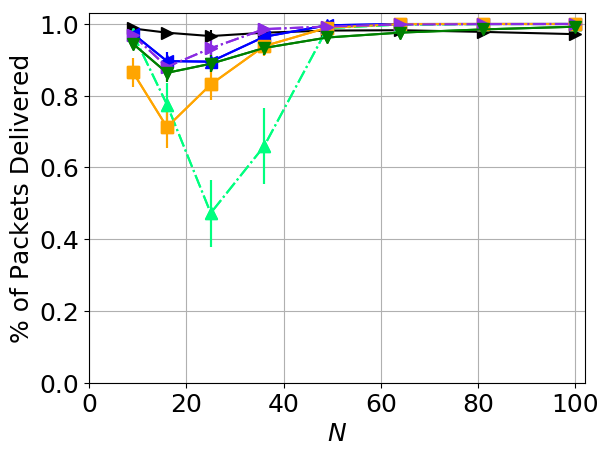}}
}}
\vspace{-0.05in}
\centerline{\hbox{
        \subfigure[Static random + high traffic]{
        \includegraphics[width=1.78in]{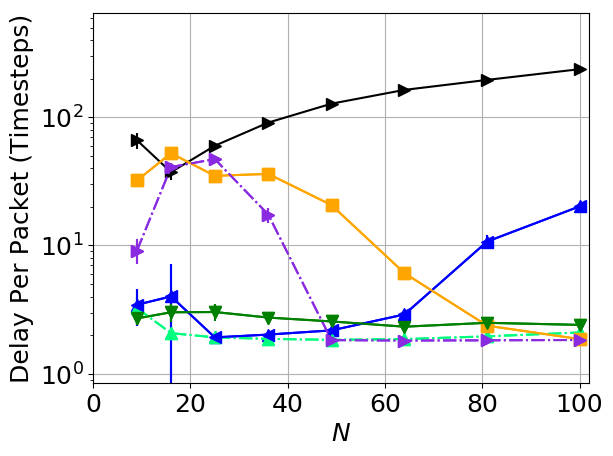}}
        \hspace{-0.18in}
        \subfigure[Delay tol. random + high traffic]{
        \includegraphics[width=1.78in]{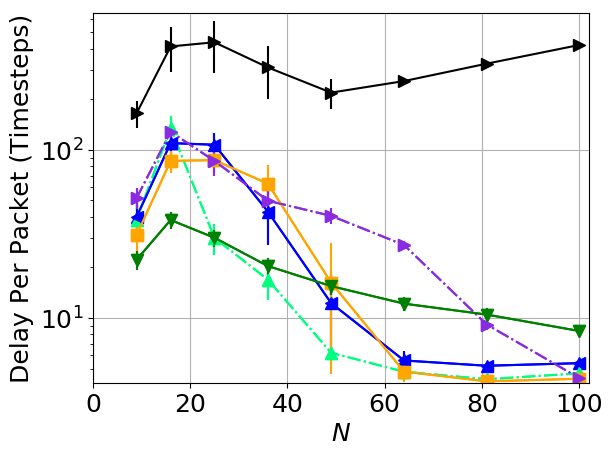}}
}}
\vspace{-0.05in}
\caption{{\small Testing generalization of DRL64 agents on the random geometric topologies; connectivity increases as $N$ increases.%
}}
\label{fig-results-traffic-cross-rand}
\vspace{-0.2in}
\end{figure}

This section shows results when training DRL64 agents on lattice or random scenarios, and then testing on just the random scenarios. This evaluates how well DRL64 agents generalize to more  diversely connected scenarios.
Fig. \ref{fig-results-traffic-cross-rand} shows results when testing on the random scenarios, for DRL64 agents trained on two lattice and two random scenarios.

Figs. \ref{fig-results-traffic-cross-rand}(a) and (c) show testing on the {\em  static random + high traffic} scenario.
As shown in Fig. \ref{fig-results-policies}(a), the static random topology has significantly higher and more variable connectivity than does the lattice, leading to many more actions to consider for each packet, as well as varying the number of neighbors over which features are computed.
Figs. \ref{fig-results-traffic-cross-rand}(a) shows that SP and DRL64 agents deliver all packets for the larger network sizes, which are more connected. For the smaller, less connected network sizes, all but the DRL64 agent trained on the delay tolerant lattice + high traffic scenario are able to deliver most packets.
Due to the decreased traffic congestion as $N$ increases, however, BP does not deliver all packets for the larger network sizes. BP also typically has significantly higher packet delay.

Figs. \ref{fig-results-traffic-cross-rand}(b) and (d) show testing on the {\em delay tolerant random + high traffic} scenario, which is disconnected for small $N$ but connected for large $N$, see Fig. \ref{fig-results-algebraic-connectivity}(a).
We now see a split in DRL64 agent performance.  Agents trained on the delay tolerant scenarios deliver all packets with the lowest delay for large $N$ but perform the worst of all strategies for small $N$; these agents have highly optimized their strategies to well-connected topologies with dynamic links, which does not generalize well to poorly connected topologies with dynamic links. Agents trained on the static scenarios also deliver all packets for large $N$ but with higher delay, while for small $N$ they perform as well as or better than SP. SP only delivers all packets for $N=100$, the most connected scenario.  While BP never delivers all packets, it delivers the most packets for small $N$ though with the highest delay, but again BP is aided by its larger queue size and ability to choose any packet in the queue.

\mysubsection{Discussion}

Our simulation results highlight the importance of training DRL agents on scenarios that are sufficiently diverse and cover the testing state space. For example, the DRL64 agent trained on the delay tolerant random + high traffic scenario (in Fig. \ref{fig-results-traffic-cross-rand}) performs exceedingly well for the $N=64$ version of the scenario on which it was trained, but does not generalize well to the smaller
versions of this scenario which are sparsely connected and highly congested.
The reason is that during training,
the $N=64$  version of the scenario lacks the occasional spikes in delay and queue length seen for the other DRL64 agents,  see Fig. \ref{fig-results-training}, so cannot generalize its learned policy to these parts of the state space during testing.

Conversely, we have results (figures not shown) for training a DRL64 agent on the static random + high traffic scenario and then testing  on all lattice scenarios. This agent generalizes well to the lattice scenarios, due to the better coverage of the state space during training.
While in this work we trained individual DRL64 agents on different scenarios, to understand the network conditions under which our approach works well, ideally,  a single DRL agent should be trained on a diverse set of scenarios that sample the space of target testing conditions.

The flexibility to choose any packet in the queue to send as with BP,
rather than only the packet at the front of the queue as with the DRL64 agents and SP,
is valuable when links are up infrequently and not all packets can be delivered. For instance, when a link is finally up, the ``best'' packet in the queue can be chosen for the link. 
Or in the case of congestion, when all packets cannot be delivered, packets that are ``easier" to deliver can be selected from the queue to be forwarded.
Incorporating this kind of flexibility into our DRL agent design would start to merge our packet-centric approach with that of a device-centric approach.

Finally, there is an interesting trade-off  between the maximum queue size $B$ and discount factor $\gamma$, since the larger $B$ is, the longer amount of time options may take, which impacts reward. We leave exploring this trade-off to future work.

\mysection{Conclusions and Future Work}
\label{sec:conclusions}
In this work, we have designed a novel distributed routing algorithm using relational deep reinforcement learning. Our algorithm generalizes to diverse  network scenarios through the use of relational features, packet-centric decisions, and extended-time actions, 
and outperforms shortest path routing and backpressure routing with respect to packets delivered and delay per packet.
There are a number of directions for future work, including extending our design to consider mobile devices and increasing flexibility in choice of packet to send.

\section*{Acknowledgemnts}
The authors acknowledge the MIT SuperCloud and Lincoln Laboratory Supercomputing Center for providing HPC and consultation resources that have contributed to the research results reported within this paper.


\bibliographystyle{IEEEtran}
\bibliography{IEEEabrv,routing}

\end{document}